\documentstyle[12pt]{article}
\begin{document}
\renewcommand{\thepage}{ }
\begin{titlepage}
\title{
\hfill
\vspace{1.5cm}
{\center Level spacing statistics of bidimensional Fermi liquids:}
{\center II. Landau fixed point and quantum chaos}}
\author{
R. M\'elin\\
{}\\
{CRTBT-CNRS, 38042 Grenoble BP 166X c\'edex France}}
\date{}
\maketitle
\begin{abstract}
\normalsize
We investigate the presence of quantum chaos in the
spectrum of the
bidimensional Fermi liquid by means of
analytical and numerical methods.
This model is integrable in a certain limit
by bosonization of the
Fermi surface. We study the effect on the
level statisticsof the momentum
cutoff $\Lambda$ present in the bidimensional
bosonization procedure.
We first analyse the level spacing statistics
in the $\Lambda$-restricted
Hilbert space in one dimension. With $g_2$
and $g_4$ interactions,
the level
statistics are found to be Poissonian at
low energies, and G.O.E.
at higher energies, for a given cut-off
$\Lambda$. In order to
study this cross-over,
a finite temperature is introduced as a
way of focussing, for a large
inverse temperature $\beta$, on the low
energy many-body states,
and driving the statistics from G.O.E.
to Poissonian.
As far as two dimensions are concerned,
we diagonalize the Fermi liquid Hamiltonian
with a small number of
orbitals.
The level spacing statistics are found to
be Poissonian in the
$\Lambda$-restricted Hilbert space,
provided the diagonal elements are
of the same order of magnitude as
the off-diagonal matrix elements of the
Hamiltonian.
\end{abstract}
\vspace{2 cm}
To be published in {\it Jour. Phys I (France)}

\end{titlepage}
\newpage
\renewcommand{\thepage}{\arabic{page}}
\setcounter{page}{1}
\baselineskip=17pt plus 0.2pt minus 0.1pt
\tableofcontents

\section{Introduction}
The ideas of quantum chaos have recently
been applied to the field
of strongly correlated electron theory
\cite{ref22} \cite{ref24}
\cite{ref25}.
These methods allow a non perturbative
description of the statistical
properties of a Hamiltonian of strongly
correlated electrons, and
may be a tool to extract some information
from finite size systems.
The aim of this article is to understand
whether the methods of quantum
chaos can shed some new light on the
problems of strongly correlated
bidimensional Fermi systems.
The question whether the bidimensional
Hubbard model is a Fermi liquid or
not is controversial. From a
theoretical point of view, Anderson
\cite{ref16} suggests that
the ground state of the bidimensional
Hubbard model is similar to
the one of the one-dimensional Luttinger
liquid, with $Z=0$ in the
thermodynamic limit, whereas Engelbrecht
and Randeria \cite{ref17}
argue that the Fermi-liquid theory is not
violated, so that this
question is controversial \cite{ref18}.
Numerical computations
lead also to controversial answers.
For instance, Dagotto et al
\cite{ref19} \cite{ref20} observe
quasiparticules in the bidimensional
t-J model, whereas Sorella \cite{ref21}
emphasizes the Luttinger liquid
behaviour.
{}From the point of view of quantum chaos,
one has to answer the
question: do level spacing statistics contain
information about
the Fermi liquid behaviour or non Fermi liquid
behaviour of the t-J model?
Before analysing the level spacing statistics
of the t-J model \cite{ref23},
we analyze in this paper models with a well
established
physical content, that is the Fermi liquid.
As we shall see, we can answer only
partially the previous question:
one can detect
integrable modes at low energy, but
quantum chaos by itself does not give
information as to whether these modes
are quasiparticule
modes or Luttinger liquid
modes. However, there is one case in
which one can conclude from
quantum chaos: in the absence of integrable
degrees of freedom
at low energy, one can conclude to the
absence of a Fermi liquid at
low energy.
Notice here the difference between
the approach of \cite{ref22} and
our point of view. In \cite{ref22},
all the energy levels of the t-J
model are analyzed on an equal footing,
whereas we shall focus essentially on
 the low energy degrees of freedom
of the Fermi liquid.

We have established in the first
article of this series \cite{ref1}, that
the level spacing statistics of the
 bidimensional Landau Hamiltonian are
poissonian. In this model, the
 quasiparticules are in a
one to one correspondence
with the non interacting gas of
spinless electron excitations, and thus
occupy orbitals labelled by the same
quantum numbers $k$ as in the case of
the gas. The quasiparticules interact
among themselves in a diagonal manner
\begin{equation}
H[\{\delta n_{\bf k}\}] =
\sum_{{\bf k}} \epsilon_{{\bf k}}
\delta n_{{\bf k}} +
\frac{1}{L^{D}}
\sum_{\langle {\bf k},{\bf k'} \rangle}
f_{{\bf k} {\bf k'}}  \delta n_{{\bf k}}
\delta n_{{\bf k'}}
{}.
\label{eq1}
\end{equation}
A numerical computation proved that the
level spacing statistics of the
Hamiltonian (\ref{eq1}) are poissonian
in two dimensions, and close to Poisson
in one dimension, with our
truncation of the Hilbert space.
The notion of level spacing statistics
seems to be relevant for quantum
fluids, and one has to distinguish
between one and two dimensions.
The link between the breakdown of the
 Fermi liquid picture in one dimension
and the level spacing statistics has
already been studied in reference
\cite{ref2}.
As far as two dimensions are concerned,
we showed in the paper I \cite{ref1},
that the Landau liquid was characterized
 by its generical integrability,
namely by Poisson level spacing statistics
 in two dimensions.
Again, the level
spacing statistics are a good tool to see
whether the Fermi system is at
the Landau fixed point or not. For the
liquid to be a Fermi liquid,
one should
be able to generate quasiparticules of
the interacting system by a switching
on procedure. In the framework of the
 Landau theory, the success of the
switching-on procedure \cite{ref3} \cite{ref10}
suggests the conservation of the number
of conserved quantities at
low energy during the switching-on
procedure, namely that the level spacing
statistics of the gas and of the interacting
liquid belong to the same
universality class. Typically, adding a non
 diagonal perturbation to
(\ref{eq1}) and obtaining a Gaussian
Orthogonal Ensemble (G.O.E.)
level spacing
statistics would mean a
departure from the Landau fixed point.
To be at the Landau fixed point,
 one must exclude strong
correlations between levels.
The corresponding generic level spacing
statistics are poissonian. However,
the case of the one dimensional Landau
Hamiltonian shows that the
statistics may not be exactly poissonian,
but close to a Poisson law.
Poisson or close to Poisson level
spacing statistics are not a sufficient
condition for the system to exhibit
quasiparticules
since one could imagine a situation in which
the same mechanism as in one dimension
for the breakdown of the
Fermi liquid holds, so that the level
 spacing statistics would remain
poissonian even in the non Fermi liquid case.
However, the adiabatic procedure, as
emphasized by Anderson \cite{ref3},
is performed within a finite characteristic
time $1/\epsilon$, and the
thermodynamic limit is taken for a finite
$\epsilon$ which is then much
larger than the typical level spacing.
So the adiabatic continuation
does not generate
true eigenstates (for this to be the case,
one should take the limit
$\epsilon \rightarrow 0$ first, before the
thermodynamic limit).
This is why the notion of adiabatic
continuation advocated to introduce
Fermi liquid is weaker than a similar
requirement for all the eigenstates
taken separately. So, this is why we
should ask whether a Fermi
liquid follows Poisson or G.O.E. statistics.

The aim of the present paper is to
study the level spacing statistics
of the Hamiltonian of spinless electrons
\begin{eqnarray}
\label{eq2}
H &=& H^{0} + H^{1}\\
\label{eq9}
H^{0} &=& \sum_{{\bf k}} \epsilon({\bf k})
 c_{{\bf k}}^{+} c_{{\bf k}}\\
H^{1} &=& \frac{1}{2 V}
\sum_{{\bf k},{\bf k'},{\bf q}} f_{{\bf k},
{\bf k'},{\bf q}}
c_{{\bf k}+{\bf q}}^{+} c_{{\bf k'} -
{\bf q}}^{+} c_{{\bf k'}} c_{{\bf k}}
,
\end{eqnarray}
where $V$ is the volume of the system.
The Hamiltonian (\ref{eq1}) describes
a Fermi liquid for times
inferior as the decay
time of the quasiparticules, and leads
 to good thermodynamical predictions
\cite{ref10}. However, the phenomenon
of decay of the quasiparticules is
not described by the Landau form
(\ref{eq1}). Since quasiparticules are true
eigenmodes of (\ref{eq1}), they have
 an infinite life-time. By contrast,
the form (\ref{eq2}) takes into account
the decay of the quasiparticules
and it is indeed possible to calculate
the decay rate of the quasiparticules
from the Hamiltonian (\ref{eq2}) \cite{ref11}.
 Even though the Hamiltonian
(\ref{eq2}) it is not diagonal,
 it can be brought
to diagonal form using some asumptions
and by bosonizing the Fermi surface
\cite{ref4}. We shall review the main
assumptions and the bosonization
of (\ref{eq2}) in the thermodynamic
limit and in the limit of a zero
curvature of the Fermi surface.
One of the ingredients of the solution
via bosonization is the existence of
 a momentum $\Lambda$ which
determines the single particule states
which participate in the formation of
bosons in the vicinity of the Fermi surface.
$\Lambda$ is a necessary ingredient,
 because of the curvature of the
Fermi surface, but is also a source
 of difficulty since the objects with
a true bosonic character in the
limit $\Lambda \rightarrow + \infty$
are no more exactly bosonic in the
limit of a finite momentum cutoff $\Lambda$.

Of course, numerical computations
cannot be performed in the thermodynamic
limit since one can only diagonalize
matrices of size 2000 by a Jacobi method.
This technical limitation imposes to
work with small systems and to impose a
 drastic cutoff $\Lambda$, so that
the Hamiltonian (\ref{eq2}) is no more
integrable in the framework of
the Hilbert space of the numerical diagonalisations.
In the case of the presence of a cutoff,
and for finite size systems,
one may thus expect
the universal Poisson level spacing
statistics to be replaced by the universal
G.O.E. level spacing statistics of the form
\begin{equation}
P(s) = \frac{\pi}{2} s
\exp{\left(- \frac{\pi s^{2}}{4} \right)}
{}.
\end{equation}
It is necessary to determine the
importance of the cutoff $\Lambda$ on the
level spacing statistics and whether
the statistics evolve towards G.O.E.
level spacing statistics as $\Lambda$
is reduced.
This has not only a numerical interest,
but also a physical one.
For instance, lattice models have a
natural cutoff $\Lambda \sim 1/a$,
where $a$ is the lattice spacing.
To answer this question, we come back to
one dimensional spinless fermionic systems,
 but with a cutoff $\Lambda$ and
study the level spacing statistics
for $g_2$ and $g_4$ interactions.
As we shall see, the level spacing
statistics are drastically affected by
the presence of the cutoff since the
bosonic superpositions of particule-hole
pairs which guarantee the integrability,
do not survive in the presence of
the cutoff. If the cutoff $\Lambda$ is
fixed and the size $L$ of the system
increases, one expects for a crossover
from a G.O.E. level
spacing statistics in the high energy
part of the spectrum to a
Poisson level spacing statistics at low energy,
which shows that finite size effects
are drastic in the presence of the cutoff.
In order to determine the crossover
scale between the Poisson regime
and the G.O.E. regime, we introduce
temperature-dependent level spacing
statistics. The statistics at finite
temperatures are found to pass from a
non integrable statistics at high
temperatures to Poisson statistics at low
temperature.

As far as two dimensionnal systems
are concerned, we carry out numerical
diagonalisations of a small system
of electrons, with a small number of
orbitals. In these conditions, the
Hamiltonian (\ref{eq2}) is non
integrable. We found Poisson level
spacing
statistics even for a small system.
We attribute this property to the fact
that the interactions generate extra
diagonal matrix elements in the
Hamiltonian which are in competition
with diagonal interaction matrix elements,
but the interaction connects only a
small number of states, leaving a lot of
zero matrix elements in the Hamiltonian. Provided
the off-diagonal matrix elements are of the same order than
the diagonal ones, the statistics
are dominated by the diagonal matrix elements.
However, by keeping only off-diagonal
interaction matrix elements, we
were able to exhibit G.O.E. level spacing statistics.

This paper is organized as follows.
 We first treat the one dimensional case,
in the Hilbert space restricted by
the momentum cut-off $\Lambda$.
The results established in one
dimension are helpful in two dimensions
since the presence of the momentum
cut-off is also a source of non
integrability. In a second step, we
 come to the bidimensional Fermi liquids,
with the study of the level spacing
statistics of the Hamiltonian
(\ref{eq2}).

\section{Level spacing statistics of 1 D
 spinless fermion systems in the
$\Lambda$-restricted Hilbert space}

\subsection{Bosonization and level spacing statistics
in the unrestricted Hilbert space (Luttinger liquid)}

Using the bosonization procedure of Haldane \cite{ref8},
we can solve the one
dimensional spinless Luttinger liquid
 with a linear dispersion relation and
$g_2$ and $g_4$ interactions. The
 Hamiltonian reads, in term of fermions
\begin{equation}
\label{eq6}
H = v_F \sum_{k,\alpha} (\alpha k - k_F)
:c_{k,\alpha}^{+} c_{k,\alpha}:
+ \frac{\pi}{L} \sum_{\alpha,\alpha'}
 \sum{q} (g_{4 q} \delta_{\alpha
\alpha'} + g_{2 q} \delta_{\alpha,
-\alpha'}) \rho_{q,\alpha}
\rho_{-q,\alpha'}
,
\end{equation}
where the label $\alpha$ indexes the
branch $\alpha$=R(ight),L(eft)$=+1,-1$.
The interaction $g_4$ describes the
diffusion of two fermions on the
same right or left
branch whereas $g_2$ describes the diffusion
of two fermions belonging
to the right and left branch.
The density operators are defined as
\begin{equation}
\rho_{q,\alpha} = \sum_{k}
:c_{k+q,\alpha}^{+} c_{k,\alpha}:
,
\end{equation}
and obey bosonic commutation relations
\begin{equation}
[\rho_{q,\alpha},\rho_{q',\alpha'}^{+}]
 = - \delta_{\alpha,\alpha'}
\delta_{q,q'} \frac{L \alpha q}{2 \pi}
,
\end{equation}
which allows the definition of boson operators
\begin{equation}
a_q^{+} = (\frac{2 \pi}{L |q|})^{1/2}
 \sum_{\alpha} \theta(\alpha q)
\rho_{q,\alpha}
,
\end{equation}
and make it possible to diagonalize the
 Hamiltonian (\ref{eq6})
via a Bogoliubov transformation:
\begin{equation}
H = E_0 + \sum_{q} \omega_q b_q^{+} b_q + \frac{\pi}{2 L}
(v_N N^{2} + v_J J^{2}),
\label{eq110}
\end{equation}
where
\begin{eqnarray}
b_q^{+} &=& \cosh{\varphi_q} a_q^{+}
 - \sinh{\varphi_q} a_{-q}\\
\tanh{2 \varphi_q} &=& -\frac{g_{2 q}}
{v_F + g_{4 q}}\\
\omega_q &=& |(v_f + g_{4 q})^{2} -
(g_{2 q})^{2}|^{1/2} |q|\\
v_S &=& \left( (v_F+g_{4 0})^{2}-
(g_{2 0})^{2} \right)^{1/2}\\
v_N &=& v_S \exp{- 2 \varphi_0}\\
v_J &=& v_S \exp{2 \varphi_0}
{}.
\end{eqnarray}
The interaction functions $g_2(q)$ and
$g_4(q)$ are supposed to tend to
a constant in the limit $q \rightarrow 0$,
 and to zero in the limit
$q \rightarrow + \infty$. Their decrease
is controlled by the impulsion
scale $2 \pi/R$, where $R$ is a given length scale.
In the thermodynamic limit
($L \gg R$), one can define an effective
low energy Hilbert space.
To do
so, we assume that $\omega_q \simeq \omega_0$
 for all the wave vectors $q$.
Of course, this approximation is only valid
if the temperature is low enough.
In this limit, the Hamiltonian (\ref{eq110})
depends only on two velocities
$v_N$ and $v_J$:
\begin{equation}
H = E_0 + \sum_{q \ne 0} (v_N v_J)^{1/2} |q| b_q^{+} b_q
+ \frac{\pi}{2 L} (v_N N^{2} + v_J J^{2})
\label{eq111}
{}.
\end{equation}
In one dimension, the link between the
breakdown of the Fermi liquid picture
and the level spacing spacing statistics
is well understood.
The absence of a Fermi surface in one dimension,
 in the presence
of long range $g_2$ and $g_4$ interactions
was known to
Dzyaloshinskii and Larkin in the 70's \cite{ref12}.
The breakdown of the Fermi liquid picture
 is already present at the level
of the static correlation functions,
and in the Hilbert space of low
energy. The usual infrared divergencies
 are governed by the $q \rightarrow 0$
limit of $g_2(q)$ and $g_4(q)$, and are
related to the orthogonality
catastrophy and the absence of a Fermi surface.
 The infrared spectrum is
described by the Hamiltonian (\ref{eq111}).
If the total charge and current
quantum numbers are given, the $q \ne 0$
excitations are bosons with a linear
dispersion relation, but with a renormalized
velocity. The departure
from the gas behaviour is measured by the
 anomalous exponents which
appear in the static Green's functions.
 The interaction energy scale
associated to this {\it static} breakdown
of the Fermi liquid is given
by \cite{ref2}
\begin{equation}
g_2^{stat} \sim v_F \left( \ln{\frac{L}
{2 \pi R}} \right)^{-1/2}
{}.
\end{equation}
However, as emphasized in \cite{ref2},
the static part of the Green's
function does not contain all the physics
of the breakdown of the
Fermi liquid theory. From a calculation
of the two points Green's function,
and from a calculation of the
switching-on pocedure, we found in
\cite{ref2} a {\it dynamical} breakdown
of the Fermi liquid picture,
controlled by the interaction scale
\begin{equation}
g_2^{dyn} \sim \frac{4 \pi v_F}
{(k L (k R)^{\alpha})^{1/2}}
,
\end{equation}
where $\alpha$ is defined by
\begin{equation}
g_{2 q} = g_{2 0}(1-(q R)^{\alpha})
{}.
\end{equation}
The static process of breakdown of the
Fermi liquid picture does not
modifie the level spacing statistics,
which remain singular, as in the case
of the gas,
since the dispersion relation of the bosons
 remain linear. By contrast,
the dynamical breakdown of the Fermi
liquid appears to be related to
a cross-over in the energy level spacing
statistics, namely from singular
level spacing statistics to generical
Poisson level spacing statistics.
As far as symetries are concerned,
the gas case and the Luttinger liquid
possess the conformal symetry. The
static breakdown of the Fermi liquid
preserves the conformal invariance.
The theory is the
antiperiodic-antiperiodic sector of
the compactified boson, with an
interaction-dependent radius of
compactification. The
dynamical breakdown of the Luttinger
liquid corresponds to a massless
loss of conformal invariance \cite{ref26}.

\subsection{Fourier transform of the
gas spectrum and temperature-dependent
level spacing statistics}
We first wish to characterize the
spectrum of the gas in the absence
of interactions, a momentum cutoff
$\Lambda$ and a finite size $L$.
The idea is to find a criterium to
detect when a boson of given
wave vector $q$ is present or not
in the system. Since $g_2$ and $g_4$
interactions are diagonal on the
basis of bosonic excitations, this tool
is a good way to characterize the
degree of integrability of the model
with a momentum cut-off.
We consider a
one branch model, and the number of quantum states
is simply the integer part
of $2 \Lambda L/2 \pi$. The spectrum is made up of
levels with an equidistant
separation $2 \pi v_F/L$. In order to characterize
the spectrum, we use
its Fourier transform
\begin{equation}
f_L(\tau) = \int_{- \infty}^{+ \infty} e^{i \omega \tau}
\sum_{\{E_i\}} \delta(\omega - E_i)
= \sum_{\{E_i\}} e^{i E_i \tau}
{}.
\end{equation}
In the thermodynamic limit,
\begin{equation}
f_{\infty}(\tau) =
\int_{- \infty}^{+\infty} e^{i \omega \tau}
\sum_{\{n_q\}} \delta(\omega-v_f
\sum_{n_q=0}^{+ \infty} q n_q)
,
\end{equation}
leading to
\begin{equation}
|f_{\infty}(\tau)| = \prod_{q>0}
\frac{1}{2 |\sin{v_f \tau q/2}|}
{}.
\end{equation}
This function presents poles which
are characteristic of
the bosonic modes for
$\tau_{n,q} =2 \pi n/v_f q$, with $n$ an integer.
If we rescale $\tau$ by a factor
$2 \pi$ and choose $v_f=1$, we obtain a
pole for $\tau_{n,q}=n/q$,
that is for every rational number.
We plotted $|f_{\infty}(\tau)|$
for 5 bosonic modes on
figure \ref{fig1}. One can easily
recognize the different bosons in the
sequence of poles. In the presence
of interactions, the poles are deplaced.
We also studied the Fourier transform
of the spectrum in the case of a finite
size system. The result is depicted
on figure \ref{fig2}. We can see that
the poles are not so well defined.
However, we can recognize the formation
of peaks which replace the poles,
and attribute well defined boson wave
vectors to some peaks. The corresponding
truncated density operators
are defined by
\begin{equation}
\rho_{\Lambda}(q) = \sum_{k} \theta_{\Lambda}(k)
\theta_{\Lambda}(k+q)
c_{k+q}^{+} c_k
\label{eq12}
,
\end{equation}
where $\theta_{\Lambda}(k)=\theta(\Lambda-|k-k_f|)$.
We can ask under which condition a
 peak corresponding to a truncated
boson $\rho_{\Lambda}(q)$ appears on
the modulus of the Fourier transform
of the spectrum $|f_L(\tau)|$. To answer
this question, we assume that the
number $n_q$ of bosonic excitations of
wave vector $q$ appearing in the
presence of a cutoff is such as
$q n_q \sim 2 \Lambda$, so that
$f_L(\tau_{1,q}) \sim 2 \Lambda / q$ for
 the peak at $\tau=\tau_{1,q}$.
A boson of wave vector $q$ is well defined
provided $f_L(\tau_{1,q}) \gg 1$,
that is if $\Lambda \gg q/2$.
For a given cut-off, the statistics are
 expected to be Poisson at
low energies, and G.O.E. at higher energies.
This is due to the fact that
truncated bosons are created essentially at low energy.
In order to test this idea, we
introduce temperature-dependent level
spacing statistics, which are
defined as follows.  If $\{E_i\}$ is the
full spectrum and $\{\epsilon_i\}$
are the energy levels after the smoothing
procedure \cite{ref13}, the
density of level spacing $P(s)$ is
\begin{equation}
\label{eq11}
P_{\beta}(s) = \left( \sum_i
\exp{\left(-\beta \frac{E_{i+1}+E_i}{2}
\right)}\right)^{-1}
\sum_i \exp{\left( - \beta \frac{E_{i+1}+E_i}{2} \right)}
\delta(s-(\epsilon_{i+1}-\epsilon_i))
{}.
\end{equation}
Moreover, for the statistics to be
comparable, one needs to scale
$P_{\beta}(s)$ and $s$ such as
$\langle P_{\beta}(s) \rangle = 1$,
and $\langle s P_{\beta}(s) \rangle =1$.
The reason why we have to rescale
 these quantities is that they were
equal to unity after the smoothing
procedure, which was carried out
in the zero temperature limit.
This property is no more valid with the
statistical weights of equation (\ref{eq11}).
For an infinite temperature ($\beta=0$),
we recover the usual level
spacing statistics. As the inverse
 temperature $\beta$ increases,
the low energy levels carry more and
more statistical weight. As we saw
from the description in terms of the
truncated bosons (\ref{eq12}),
the spectrum is expected to be
integrable at low energies,
so that the level
spacing statistics should evolve
from G.O.E. statistics, or at least
intermediate statistics
(that is, with $0 < P(0) < 1$), to Poisson
level spacing statistics as $\beta$ increases.

\subsection{Level spacing statistics
with $g_2$ interactions
in the $\Lambda$-restricted Hilbert space}
The numerical method to compute the
level spacing statistics of two
branch models with a momentum cutoff
$\Lambda$ and $g_2$ interactions
consists in tabulating the states of
the Hilbert space by generating all the
different fillings of the two branches,
the number of
particules on each branch
being kept constant. The second step
consists in computing all the matrix
elements of the hamiltonian and
diagonalising the Hamiltonian by the use
of the Jacobi method. The interactions are of the form
\begin{equation}
g_{2 q}=g_{2 0} \exp{- \left( \frac{q L}{2 \pi R} \right) }
{}.
\end{equation}
The scale of the interactions $R$ is
 chosen equal to $\Lambda$, so that all
the matrix elements of the interaction
are important. The evolution of
the spectrum as a function of $g_{2 0}$
is plotted on figure \ref{fig3}.
This figure is to be compared with the
figure 2 of reference \cite{ref2},
where we have plotted the evolution of
the energy levels as a function
of $g_{2 0}$ but in the absence of cutoff.
The obvious difference between
the two plots is the presence of level
repulsion with a cutoff $\Lambda$, and
the existence of level crossings in the absence of cutoff.
The level spacing statistics corresponding to
a spectrum analog to the one
of figure
\ref{fig3} are plotted on figure \ref{fig4}.
They are in good agreement
with G.O.E. level spacing statistics.
In our computation, $\Lambda L/2 \pi =3$,
so that the criterium $\Lambda \gg 2 \pi/L$
is not verified.
We could not go to higher values of
$\Lambda L/2 \pi$ because increasing
the ratio $\Lambda L/2 \pi$ increases
the size of the Hilbert space and
numerical diagonalizations are no more possible.
We conclude that the statistics depend drastically
on the length of the system,
if the momentum cutoff is fixed.
 If we apply the ideas of
temperature-dependent statistics of
level spacings to the system, we find
that the statistics of level spacings
are driven from a G.O.E. law at
$\beta=0$ to a Poisson law as $\beta$
increases, as depicted on figure
\ref{fig4bis}. This evolution reveals
the fact that "nearly integrable"
degrees of freedom exist at low energy.
 These bosonic degrees of freedom
are exactly integrable in the absence of
 cut-off. In term of classical
trajectories (provided one is able to find
 a classical phase space
for the Fermi liquid !), this situation
corresponds to the existence of
conserved torii at low energy, which
transform into chaotic trajectories
as the energy increases. The cross-over
temperature scale shall be
derived later.

\subsection{Level spacing statistics with $g_4$ interactions
in the $\Lambda$-restricted Hilbert space}
We now consider the case of $g_4$ interactions.
The interaction Hamiltonian is
\begin{equation}
H^{1} = \sum_{q\ne 0} \sum_{k,k'\ne k+q}
g_{4 q} c_{k+q}^{+} c_{k'-q}^{+} c_{k'} c_k
{}.
\end{equation}
The interest of the $g_4$ term is that we can
use only a one branch model, and we can
 reach higher values of the ratio
$\Lambda L/2 \pi$ without increasing the
 size of the Hilbert space.
We could reach $\Lambda L/2 \pi = 12$ in
a sector of total momentum
$P = 24.2 \pi/L$.
$P$ is the impulsion with respect to the
fundamental. As long as
$P \le \Lambda L/\pi$, one has generated
the complete Hilbert space in the
absence of interactions. This fact motivates
the choice $P= \Lambda L /\pi$.
The condition $\Lambda L/\pi = 24 \gg 1$
is respected.
However, we did not find Poisson statistics,
but statistics which are
intermediate between a Poisson law and
G.O.E. statistics. Namely, the value
of the density of normalized zero
crossings is not $1$ but $0.5$.
The statistics are plotted on
figure \ref{fig5}. This computation gives an
idea of the extension of the
crossover as a function of $\Lambda L/2 \pi$,
since we obtain intermediate
statistics for $\Lambda L/2 \pi=24$.
We computed also the temperature-dependent
 level spacing statistics.
We found a cross-over to Poisson
statistics as the inverse temperature
increases, as plotted on figure \ref{fig5bis}.
This cross-over is much more
rapid than in the case of the
$g_2$ interaction. However, the
values of the cutoff
in the $g_2$ case and the $g_4$
case are not comparable.

\subsection{Cross-over scales}
In order to get an idea of the temperature
cross-over between the G.O.E. and
Poisson regime, we look for the
energy scale $k_B T^{*}$ below which
the energy levels are decorrelated.
To do so, we make the approximation
that the only effect of interactions
is to compress the spectrum by an amount
$((1 + g_4/v_f)^{2}-(g_2/v_f)^{2}))^{1/2}$,
where $g_2$ and $g_4$ are
typical energy interactions. This
compression of the spectrum
means that the effective mass increases if
$g_2$ is large compared to $g_4$.
This approximation corresponds to
taking local interactions in the real space.
Then, we see that the
integrable modes under request are such as
\begin{equation}
v_f |q| \le \left( \left(1+\frac{g_4}{v_f}\right)^{2}
- \left(\frac{g_2}{v_f}\right)^{2} \right)^{-1/2}
k_B T
{}.
\end{equation}
These modes are integrable provided
they lead to well-defined bosons.
We say that a boson is well-defined
provided a sufficient number of
particule-hole excitations enter into
the summation (\ref{eq12}), that is
if $2 \Lambda - |q| \ge  2 \pi \alpha/L$,
where $\alpha$ is a dimensionalless
coefficient, which counts the number
of non-zero terms in (\ref{eq12}), and
is thus a measure of integrability.
We get that
\begin{equation}
k_B T \simeq (2 \Lambda - 2 \pi \alpha/L)
 \left( (1+ g_4/v_f )^{2} - (g_2/v_f)^{2} \right) ^{1/2}
\end{equation}
The energy scale $k_B T^{*}$ thus
 increases as the cut-off $\Lambda$ increases,
and decreases if the interaction
strength $g_2$ increases.

\section{Level spacing statistics
 of a 2 D spinless fermion system in the
$\Lambda$-restricted Hilbert space}
\subsection{Bosonization of the bidimensional Fermi surface}
The bosonization of the Fermi
liquid \cite{ref14}, \cite{ref6}, \cite{ref7}
involves a covering of the Fermi
surface by spheres of radius $\Lambda$.
Each small sphere is labelled by an
integer $\alpha$ and one assumes that
the Fermi surface is flat at the scale $\Lambda$.
Assuming that $\Lambda \gg 2 \pi / L$,
the commutation relations of the
operators
\begin{equation}
\label{eq8}
\rho_{q,\alpha} = \sum_{{\bf q}}
\theta_{\alpha}({\bf k} + {\bf q}/2)
\theta_{\alpha}({\bf k} - {\bf q}/2)
 n_{{\bf q}}({\bf k})
,
\end{equation}
with
\begin{equation}
n_{{\bf q}}({\bf k}) = c_{{\bf k} -
{\bf q}/2}^{+} c_{{\bf k} + {\bf q}/2}
,
\end{equation}
must include a Schwinger term, leading
to the commutation relation
\begin{equation}
\label{eq10}
[\rho_{{\bf q},\alpha},\rho_{{\bf q}',
\alpha'}^{+}]=\delta_{\alpha,\alpha'}
\delta_{{\bf q},{\bf q'}}
\sum_k C_{\alpha}({\bf k},{\bf q})
(n^{0}_{{\bf k}+\frac{{\bf q}}{2}} -
n^{0}_{{\bf k}-\frac{{\bf q}}{2}})
,
\end{equation}
where the constraint $C_{\alpha}$ is
\begin{equation}
C_{\alpha}({\bf k},{\bf q}) =
\theta_{\alpha}({\bf k}+\frac{{\bf q}}{2})
\theta_{\alpha}({\bf k}-\frac{{\bf q}}{2})
\theta_{\alpha}({\bf k}'+\frac{{\bf q}'}{2})
\theta_{\alpha}({\bf k}'-\frac{{\bf q}'}{2})
{}.
\end{equation}
To simplify the expression (\ref{eq10}),
 one makes the assumption of
a flat Fermi surface in each sphere
 of radius $\Lambda$.
In order to specify this condition,
one imposes that no bosonic excitations
 with an angle $\theta > \pi/2$ exist,
where $\theta$ denotes the angle
between ${\bf k_f}$ and ${\bf q}$.
The maximal angle theta is such as
\begin{equation}
\tan{\left( \theta- \frac{\pi}{2} \right)}
 \le \frac{2 \pi}{L \Lambda},
\end{equation}
which leads to
\begin{equation}
\Lambda \le (\frac{4 \pi k_f}{L})^{1/2}
\label{eq4}
{}.
\end{equation}
The equation (\ref{eq4}) means that the
 curvature is negligeable in a sphere
of radius $\Lambda$, and is compatible
 with the condition
\begin{equation}
\Lambda \gg 2 \pi/L
\label{eq5}
\end{equation}
provided the linear length is large
enough: $L \gg \pi/k_f$.
If $\Lambda \ll 2 \pi/L$ and $|{\bf q}|
 \ll \Lambda$, the leading term in
(\ref{eq10}) is of the form
$[\rho_{{\bf q},\alpha},\rho_{{\bf q}',
\alpha'}^{+}]=\delta_{\alpha,\alpha'}
\delta_{{\bf q},{\bf q'}} V a ({\bf q}.
{\bf n_{\alpha}})$.
We evaluate the corrections under the
 conditions $\Lambda \gg 2 \pi/L$
and $|{\bf q}| \ll \Lambda$. Under
these assumptions, the leading term
represents the number of states in
the parallelogram of figure \ref{fig0}.
To obtain the corrections, one has
to substract the
number of states contained in the
small shaded triangle.
The number of states to be removed
is approximately equal to
\begin{equation}
\frac{1}{2} \left(\frac{L}{2 \pi}
\right)^{2} |{\bf q}|^{2}
\cos{\theta} \sin{\theta} = V a
({\bf q}.{\bf n}_{\alpha}).
\frac{|{\bf q} \wedge {\bf n}_{\alpha}|}{2 k_f}
{}.
\end{equation}
The commutation relations are thus of the form
\begin{equation}
[\rho_{{\bf q},\alpha},\rho_{{\bf q}',
\alpha'}^{+}]=\delta_{\alpha,\alpha'}
\delta_{{\bf q},{\bf q'}} V a ({\bf q}.
{\bf n_{\alpha}})
\{1 + O(\frac{|{\bf q}
\wedge{\bf n}_{\alpha}|}{k_f})\}
\label{eq3}
{}.
\end{equation}
Following reference \cite{ref7},
we define
\begin{equation}
a_{{\bf q}}({\bf k_f}) =
\sum_{{\bf k}}
\phi_{\Lambda}(|{\bf k} - {\bf k_f}|)
[ n_{\bf q}({\bf k}) \theta({\bf q}.{\bf v_k}) +
n_{\bf -q}({\bf k}) \theta(- {\bf q}.{\bf v_k})],
\end{equation}
and
\begin{equation}
b_{{\bf q}}({\bf k_f}) =
(N_{\Lambda} V |{\bf q}.{\bf v_k}|)^{-1/2}
a_{{\bf q}}({\bf k_f}),
\end{equation}
where $\phi_{\Lambda}$ is a
smearing function such as
$\phi_{\Lambda} \rightarrow \delta_{{\bf k}, {\bf k_f}}$
if $\Lambda \rightarrow 0$,
and $N_{\Lambda}$ is the local density of states:
\begin{equation}
N_{\Lambda} = \frac{1}{V}
\sum_{{\bf k}} |\phi_{\Lambda}(|{\bf k}-
{\bf k_f}|)^{2} \delta(\mu-\epsilon_{\bf k})
{}.
\end{equation}
Provided the curvature of the Fermi
surface is negligeable, that is provided
condition (\ref{eq4}) is satisfied,
one can bosonize the Hamiltonian
(\ref{eq2}) to obtain
\begin{equation}
H = \sum_{{\bf k_f}} \sum_{{\bf q},
{\bf q}.{\bf v_k}>0}
|{\bf q}.{\bf v_k}| b_{{\bf q}}^{+}({\bf k_f})
b_{{\bf q}}({\bf k_f})
+ \frac{1}{2 V} \sum_{{\bf k},{\bf k'},
{\bf q}} f_{{\bf k},{\bf k'},{\bf q}}
n_{-{\bf q}}({\bf k}) n_{{\bf q}}({\bf k'})
{}.
\end{equation}
Since the interactions are quadratic
in the boson operators,
one can diagonalize
them via a generalized Bogoliubov transformation.
However, in numerical computations, one
 can only treat
a Hilbert space of
size $2000$. This means that the cutoff $\Lambda$ must
be reduced drastically,
as well as the Fermi wave vector $k_f$. In particular,
the conditions
(\ref{eq4}) and (\ref{eq5}) are no longer valid and one
cannot diagonalize the
Hamiltonian as described above.
Two scenarios are candidates for the
appearance of chaos in the spectrum
of the bidimensional Fermi liquid.
The first scenario is the effect of
curvature. In the limit $L \rightarrow + \infty$,
and for a cut-off independent
of $L$, the condition (\ref{eq4}) is no more valid,
and one cannot solve
the model by bosonization. We shall not study this
effect in the present paper.
The second scenario comes from the fact that,
even with a flat Fermi surface,
the condition (\ref{eq5}) may be violated,
so that the system is no more
integrable by bosonization. We shall
study the last type of effect in the
rest of the paper.

\subsection{Level spacing statistics of a
2 D spinless fermion system in the
$\Lambda$-restricted Hilbert space}
We now turn to the bidimensional case in
the presence of a momentum cutoff.
The Fermi sea in its fundamental state
is pictured on the insert of
figure \ref{fig7}.
We treated a Fermi sea of 5 electrons
 for a total of 29 available
quantum states.
The Hamiltonian is given by the
 expression (\ref{eq2}). The interaction
term can be split into two terms as follows
\begin{eqnarray}
H^{1} &=& H^{1}_0 + H^{1}_1\\
\label{eq7}
H^{1}_0 &=& - \frac{1}{V} \sum_{{\bf k},
{\bf q}}
f_{{\bf k},{\bf k}+{\bf q},{\bf q}}
n_{{\bf k}+{\bf q}} n_{{\bf k}}\\
H^{1}_1 &=& \frac{1}{V} \sum_{q \ne 0}
\sum_{{\bf k},{\bf k'}\ne {\bf k}+
{\bf q}}
f_{{\bf k},{\bf k'},{\bf q}} c_{{\bf k}+{\bf q}}^{+}
c_{{\bf k'}-{\bf q}}^{+} c_{{\bf k'}} c_{{\bf k}}\\
\end{eqnarray}
The term $H^{1}_{0}$ is of the same
nature as the diagonal Landau interaction
between quasiparticules. The only
difference is that $\delta n_k$
represents occupation numbers of
renormalized quasiparticules in the Landau
theory, whereas $n_k$ is the number
operator of bare fermions. The
Hamiltonian made up of the kinetic
term $H^{0}$ (\ref{eq9}) plus the
term $H^{1}_0$ (\ref{eq7}) has already
been studied in reference
\cite{ref1} and leads to Poisson
statistics in two dimensions.
In the limit in which the bidimensional
bosonization procedure is
applicable, it has been established in
\cite{ref7} that the Hamiltonian
$H+H^{1}_0+H^{1}_1$ is integrable,
namely that the effect of
$H^{1}_0+H^{1}_1$ is to renormalize
the free theory without breaking the
integrability. We question whether
this property is still valid for a
system of electrons such as the one
drawn on the insert of figure \ref{fig7}.
To see this, we diagonalize the
Hamiltonian $H^{0}+H^{1}_0+H^{1}_1$ for
the electron system of the insert of figure \ref{fig7},
in a sector of fixed total
momentum.  As expected from
bosonization \cite{ref7} and from
R.P.A. theory \cite{ref9},
we obtain a collective bound state which detaches
itself from the particule-hole continuum.
 The energy of the collective
mode is greater as the continuum energy
 for repulsive interactions,
and lower for attractive interactions.
Because of the small value
of the momentum cutoff $\Lambda$, the
 energy width of the continuum
is small compared to the energy of the
 bound state, so that the bound
state renormalizes the level spacing
statistics of the continuum towards
small values, since the mean value of
the level spacing distributions is
rescaled to unity. If we suppress the
bound state from the spectrum,
we obtain level spacing statistics in
good agreement with Poisson level
spacing statistics, as shown on figure \ref{fig7}.
It shows that whereas the cutoff
$\Lambda$ is very small and the number of
electrons small too,
the result we obtain is consistent
with the result predicted
in the limit in which the bosonization
procedure is valid, with a
large cutoff compared to $2 \pi/L$.
However, there is no contradiction with
what has been done in one dimension.
If we assume that in two dimensions the
statistics are poissonian provided
a sufficient number of truncated bosons
(\ref{eq8}) have a sufficient number
of non zero terms in their linear combination,
we obtain the same criterium as in one
 dimension, namely that the statistics
are poissonian provided
$\Lambda \gg 2 \pi/L$. Using the cross-over scale
we obtained from the diagonalization of
 $g_4$ interactions, we reach the
conclusion that the level spacing
statistics of the Hamiltonian
$H^{0}+H^{1}_2$ should be of the G.O.E. type.
This is indeed the case, as
shown on figure \ref{fig8}.
We attribute the fact that the level
spacing statistics of
$H^{0}+H^{1}_0+H^{1}_1$ are poissonian
to the presence of only a small
number of off-diagonal terms, which are
 not numerous enough to change
significatively the statistics.

\section{Conclusion}
In this paper, we have studied the effect
of the momentum cutoff
on the level spacing statistics of
interacting Fermi systems.
We found that the
presence of the cutoff could change
drastically the level spacing
statistics of a finite size system,
namely to drive
the level spacing statistics
from a poissonian shape to G.O.E. statistics.
Using temperature-dependant level spacing statistics
and one dimensional models, we have shown that the
system of electrons in the presence of
a cutoff is integrable at low energy.
As far as $g_2$ interactions are
 concerned,
we obtained G.O.E. level spacing
statistics, which evolve to a Poisson
level spacing statitics as the
temperature decreases from $+ \infty$.
In the $g_4$ case, we could reach higher
 values of the cutoff since the
two branches of the dispersion relation
 are decoupled.
The infinite temperature level spacing
 statistics were intermediate
between Poisson and G.O.E., and were
driven to a Poisson shape as the
temperature decreases. The main feature
of a system of spinless correlated
fermions in the presence of a momentum
 cutoff in the $k$-space is that,
inspite of the loss of integrability
there subsists nearly uncorrelated
levels at low energy. This conclusion
is analogous to the conclusions of
reference \cite{ref15} for another
model. In two dimensions,
one has to distinguish between two cases.
The first case corresponds to
off-diagonal interactions only.
In this case,
the level spacing statistics are G.O.E.
 statistics. It is clear that in this
case, no switching on procedure can
connect the excitations of the
gas and the excitations of the
interacting system, since the nature
of the two spectra is different,
which means that the system is not at
the Landau fixed point. In the second case,
diagonal interactions of the Landau
type are taken into account.
The Poisson level spacing statistics are
restored because the off-diagonal
matrix elements are not numerous enough,
whereas their amplitude is comparable
to the amplitude of the diagonal
matrix elements. The Fermi liquid
 behaviour of the Hamiltonian in the
presence of diagonal interactions
is thus not destroyed.
This result is to be compared with
the fact that the Hamiltonian
is diagonal in the limit in which the
 bidimensional bosonization procedure
is valid. In spite of the small number
of fermions and the small number of
orbitals in our numerical computations,
we obtain a result in good
agreement with the bosonization of the
Fermi surface theory.
It should be stressed that the limit in which
the bosonization theory is valid
is not the thermodynamic limit in
 which $\Lambda$ is fixed and
$L \rightarrow + \infty$, because
of the constraint (\ref{eq4}).
The problem of the thermodynamic
limit, with the condition (\ref{eq4})
violated remains open.

The finite temperature level spacing
statistics seem to be an appealing
tool for the study of the spectrum of
 strongly correlated electronic
systems. From a finite size study, one
can characterize the integrability
at low energy, and the cross-over
temperature measures how far the low
energy degrees of freedom are from being
integrable. However, it is clear
that the level spacing statistics retain
 only information about the
the symetries, and give no information
 about the decay of the correlation
functions, so that, in principle, we
 cannot solve entirely the problem
whether the bidimensional t-J model is
a Fermi liquid or not.
Nonetheless, we can characterize the
 degree of integrability of the
low energy degrees of freedom.
A forthcoming paper shall be devoted to the case
of finite size t-J models
\cite{ref23}.

The author acknowledges B. Dou\c{c}o
 for underlying the role of the
momentum cutoff, and J.C. Angl\`es
d'Auriac for help with algorithms.

\newpage
\renewcommand\textfraction{0}
\renewcommand\floatpagefraction{0}
\noindent {\bf Figure captions}

\begin{figure}[h]
\caption{}
\label{fig1}
Modulus of the Fourier transform of the
spectrum in the thermodynamic limit
$|f_{\infty}(\tau)|$
for 5 bosons. Each pole corresponds to
one or several definite boson
excitations. We have chosen $v_f=1$ and
$L=2 \pi$. A given boson
$q >0$ generates a sequence of poles at
$\tau_{n,q}=2 \pi n/ q$. The poles are
denoted by their corresponding rational
number $n/q$.
\end{figure}

\begin{figure}[h]
\caption{}
\label{fig2}
Finite size effects of the
modulus of the Fourier transform of the
 spectrum $|f_L(\tau)|$.
We have chosen $v_f=1$ and $L=2 \pi$,
$\Lambda=11$.
As shown on the figure, we can
attribute boson quantum numbers to some peaks.
We recognize the fractions $1/5$,
$1/4$, $1/3$,$2/5$,$1/2$,$3/5$,$2/3$,$3/4$,
$4/5$.
\end{figure}

\begin{figure}[h]
\caption{}
\label{fig3}
Evolution of the energy levels in the
presence of $g_2$ interactions with
a momentum cutoff. The parameters of
the model are $v_f=1$ and $L=2 \pi$.
The cutoff is chosen equal to 3, so
that we have 3 fermions on each branch,
for 6 available quantum states. The
interactions are of the form
$g_{2 q}=g_{2 0} \exp{(-q/R)}$. with $R=6$.
The momentum sector is 1, leading
to 45 states in the Hilbert space.
Because of the particule-hole symmetry,
the spectrum is symmetric, but with no level crossings.
\end{figure}

\begin{figure}[h]
\caption{}
\label{fig4}
Level spacing statistics of 1D spinless
fermions with $g_2$ interactions
and a cutoff. The parameters are $v_f=1$,
$L=2 \pi$. In order to eliminate
the particule-hole  symmetry present on figure \ref{fig3},
we chose a non symmetric cutoff
for particules and holes: the number
of fermions on the right branch is 4,
and the number of quantum states is 9.
On the left branch, the number of
fermions is 5 for 9 quantum states.
The Hilbert space contains 1052
states in the sector of momentum 1.
The level spacing statistics are found
to be well fitted by G.O.E. statistics.
\end{figure}

\begin{figure}[h]
\caption{}
\label{fig4bis}
Level spacing statistics of 1D spinless
fermions with $g_2$ interactions
and a cutoff at finite temperature.
The parameters are $v_f=1$, $L=2 \pi$.
In order to eliminate
the symmetry present on figure \ref{fig3},
 we chose a non symmetric cutoff
for particules and holes: the number of
fermions on the right branch is 4,
and the number of quantum states is 9.
 On the left branch, the number of
fermions is 5 for 9 quantum states.
The Hilbert space contains 1052
states in the sector of momentum 1.
The level spacing statistics are
plotted for inverse temperature equal
to $\beta=0, 0.1, 0.5$
A cross-over is found from G.O.E.
statistics as $\beta=0$ to
Poisson statistics as $\beta$ increases.
\end{figure}

\begin{figure}[h]
\caption{}
\label{fig5}
Level spacing statistics of 1 D spinless
fermions with $g_4$ interactions
and with a momentum cutoff. The parameters
are chosen such as $v_F=1$
and $L=2 \pi$. The value of the momentum
cutoff $\Lambda$ is 12 and the total
momentum is 24. The Hilbert space contains
 1185 states.
The statistics is found to be intermediate
between a G.O.E. statistics and a
Poisson statistics.
\end{figure}

\begin{figure}[h]
\caption{}
\label{fig5bis}
Level spacing statistics of 1 D spinless
 fermions with $g_4$ interactions
and with a momentum cutoff. The parameters
are chosen such as $v_F=1$
and $L=2 \pi$. The value of the momentum cutoff
$\Lambda$ is 12 and the total
momentum is 24. The Hilbert space contains
 1185 states.
The statistics are plotted for the inverse
temperature equal to
$\beta=0$ and $\beta=0.3$. The statistics
for $\beta=0.3$ is close to
the Poisson statistics.
\end{figure}

\begin{figure}[h]
\caption{}
\label{fig0}
Representation of a sphere at the Fermi surface.
 The commutation relation
(\ref{eq10}) is proportional to the number
of states contained in the
intersection of the parallelogram and the
sphere $n$. The leading order
term takes into account all the states in
the parallelogram.
\end{figure}

\begin{figure}[h]
\caption{}
\label{fig7}
Level spacing statistics of the Hamiltonian
$H^{0}+H^{1}_0+H^{1}_1$.
The level spacing statistics are
 poissonian
if one excludes the bound state,
and renormalized towards $s=0$ in the
presence of the bound state. The
 linear size of the box is $2 \pi$.
The kinetic
term $\epsilon(k)$ is quadratic,
with a mass equal to 3.
The interactions
have the form $f_{k,k',q}=f^{0}
\exp{(-|q|/R)}$, with $R=6$.
The insert represents the Fermi
sea which was used. The Hilbert space
contains 1042 states in the sector
of momentum $P = (2 \pi/L) (2,1)$.
\end{figure}

\begin{figure}[h]
\caption{}
\label{fig8}
Level spacing statistics of the
Hamiltonian $H^{0}+H^{1}_1$ with the
bound state excluded. The number
of zero level
spacing is equal to 1114 , leaving
only 186 non zero level spacings
for the statistics,
which explains the important fluctuations
of the level spacing statistics
which agree with the G.O.E. shape.
The insert represents the Fermi sea which
was used. The Hilbert space
contains 1042 states in the sector of
momentum $P = (2 \pi/L) (2,1)$.
\end{figure}


\newpage
\begin{thebibliography}{99}
\bibitem{ref22} G. Montambaux, D. Poilblanc,
J. Bellisard and C. Sire, Phys.
Rev. Lett. \underline{70}, 497 (1993).
\bibitem{ref24} T.C. Hsu and J.C.
 Angl\`es d'Auriac, Phys. Rev. B
\underline{47}, 14291 (1993).
\bibitem{ref25} D. Poilblanc,
T. Ziman, J. Bellissard, F. Mila and
G. Montambaux, Europhys. Lett.
\underline{22}, 537 (1993).
\bibitem{ref16} P.W. Anderson,
 Phys. Rev. Lett. \underline{65}, 2306 (1990).
\bibitem{ref17} J.R. Engelbrecht
and M. Randeria, Phys. Rev. Lett.
\underline{65}, 1032 (1990).
\bibitem{ref18} P.W. Anderson; J.R.
Engelbrecht and M. Randeria,
Phys. Rev. Lett. \underline{66}, 3225 (1991).
\bibitem{ref19} E. Dagotto, A.
Nazarenko and M. Boninsegni, Phys. Rev. Lett.
\underline{73}, 728 (1994).
\bibitem{ref20} A. Morea, S. Haas,
A. Sandwik and E. Dagotto, preprint (1994).
\bibitem{ref21} Sorella, preprint cond-mat/9308001.
\bibitem{ref23} R. M\'elin, in preparation.
\bibitem{ref1} R. M\'elin, J. Phys.
I France \underline{5}, 159(1995).
\bibitem{ref26} P. Degiovanni and R. M\'elin, unpublished.
\bibitem{ref2} R. M\'elin,
B. Dou\c{c}ot and P. Butaud, J. Phys. I France
\underline{4}, 737 (1994).
\bibitem{ref3} Anderson,
Basic Notions of Condensed Matter Physics, Frontiers
in Physics,
The Benjamin/Cummings Publishing Company (1984).
\bibitem{ref10} D. Pines and P. Nozieres,
 The Theory of Quantum Liquids,
vol 1, Addison-Wesley Publishing Company (1966,1989).
\bibitem{ref11} A. Houghton, H.J. Kwon
 and J. B. Marston, preprint (1993).
\bibitem{ref12} I. E. Dzyaloshinskii
and A.I. Larkin Sov. Phys.
JETP \underline{38}, 202 (1974)
\bibitem{ref4} Castro Neto and E. Fradkin,
to be published.
\bibitem{ref14} F.D.M. Haldane, unpublished.
\bibitem{ref6} A. H. Castro Neto and
E. H. Fradkin, Phys. Rev. Lett.
\underline{72}, 1393 (1994).
\bibitem{ref7} A. H. Castro Neto and
E. H. Fradkin, to be published.
\bibitem{ref8} F. D. M. Haldane,
J. Phys. C \underline{14},1285 (1981).
\bibitem{ref13} M. C. Gutzwiller,
Chaos in Classical and Quantum Mechanics,
Springer Verlag (1990).
\bibitem{ref9} J.W. Negele and H.
Orland, Quantum Many Particule Systems,
Frontiers in Physics (Vol. 68), Addison-Wesley
Publishing Compagny, Inc (1987).
\bibitem{ref15} M. Di Stasio and X.
 Zotos, to be published.
\end{thebibliography}
\end{document}